\documentclass[aps,prd, 
amsmath,amssymb,showpacs,nofootinbib]{revtex4}
\usepackage{graphicx}

\newcommand{\refeq}[1]{(Eq.\,\ref{eq-#1})}
\newcommand{\fig}[4]{\begin{figure}[htbp]\centering\includegraphics[width=#3\textwidth]{#1}\caption{#2}\label{fig-#1}\end{figure}}
\newcommand{\refig}[1]{Fig.\ \ref{fig-#1}}

\newtheorem{theorem}{Theorem}

\usepackage[usenames,dvipsnames]{color}
\usepackage{hyperref}
\usepackage{color}

\def\lsim{\, \rlap{$<$}{\lower 1.1ex\hbox{$\sim$}}\,}

\newcommand{\void}[1]{}

\begin{document}

\title{Dualities between Scale Invariant and Magnitude Invariant Perturbation Spectra 
in Inflationary/Bouncing Cosmos
}

\author{Changhong~Li,}
\email[]{chellifegood@gmail.com}
\author{Yeuk-Kwan~E.~Cheung}
\email[]{cheung@nju.edu.cn \& cheung.edna@gmail.com}
\affiliation{Department of Physics, Nanjing University, 22 Hankou Road, Nanjing, China 210093}

\begin{abstract}
We study the cosmological perturbation spectra using the dynamical equations of gauge invariant perturbations with a generalized blue/red-shift term, $mH\dot\chi$. 
Combined with the power-law index of cosmological background, $\nu$, we construct a parameter space, $(\nu, m)$, to classify all possible perturbation spectra.
The magnitude-invariant power spectra occupy a two dimensional region in the $(\nu, m)$ space. 
We find two groups of scale-invariant solutions, one of which are also magnitude invariant while the other not, in the expanding and contracting phases of cosmological evolution.
We explore the implications of magnitude invariance to scale-invariance of an underlying spectrum, and unveil a novel duality between the scale-invariant solutions and the boundary of magnitude-invariant solutions: under two consecutive duality transformations, the scale-invariant solutions are mapped onto the boundary of magnitude invariant region.
The physical origin of such a duality is yet to be studied.
We present cosmological applications of our observations to a de-Sitter universe, an  Ekpyrotic background, and a matter-dominated contraction.
For the first two cases previously known results are re-derived; we also find for a matter-dominated contraction a solution with both scale-invariance as well as magnitude-invariance--
in contrast to Wands's scale-invariant but magnitude-variant spectrum.

\end{abstract}
\maketitle

\section{Introduction and Summary of Results}
\label{Intro}

The single field slow-roll inflation model has been hailed a paradigm for its much lauded rendition of scale-invariant power spectrum that shows up in an array of modern cosmological observations.
Generating scale-invariant power spectrum, however, has been shown not a monopoly of the slow-roll inflation scenario due to recent progress in model building.
It is also possible to construct which generate scale-invariant power spectrum of cosmological perturbations.

In search of scale invariant solutions for the power spectrum of cosmological perturbations, compelling alternative cosmological models have been constructed  and the possibilities of generating  scale-invariant power spectrum investigated. 
For example,  a contracting matter-dominated background was discovered  by Wands~\cite{Wands:1998yp} (see also, Finelli and Brandenberger~\cite{Finelli:2001sr}) to generate a scale-invariant spectrum.  Khoury et al~\cite {Khoury:2001zk}, by neglecting gravity back reaction in the dynamics of the fluctuations, found a scale-invariant power spectrum for an Ekpyrotic-type background (slowly  contracting or expanding)~\cite{Khoury:2001wf}. 
If, however, gravity is taken into account, the spectrum of perturbations in an Ekpyrotic background with a single canonical scalar field becomes blue~\cite{Creminelli:2004jg}. 
On the other hand, possible dualities of  power spectra between backgrounds with seemingly different equations of state--in a particular gauge--have also been studied~\cite{Boyle:2004gv}.
Power spectra with scale-invariance are more readily found for Ekpyrotic-like backgrounds once the long-held assumption that the equation of state of backgrounds be nearly constant is relaxed~\cite{Khoury:2009my}. 
In this paper we shall, nevertheless, assume the constancy of  the equation of state of the cosmological  backgrounds  during the period in which the primordial power spectra are generated.

The  brilliant example  discovered by Wands~\cite{Wands:1998yp}--scale-invariant (with $k$-independence)
power spectrum of cosmological perturbations generated during a matter-dominated contraction--lends hope to obtaining scale-invariant power spectrum in more general backgrounds other than de-Sitter.
However the example given by Wands is not truly scale-invariant because of its time dependence: the amplitude of power spectrum increases on the effective horizon.
The amplitude of each mode in the power spectrum depends on the moment at which the associated mode exits the effective horizon. The moment of exit is, in turn,  determined by the mode's $k$ value.
Altogether an implicit k-dependence creeps into the value of the Hubble parameter as the  mode exits its horizon, $H_*(k)$, rendering
the power spectrum as a whole  non-scale-invariant.
A brief discussion concerning Wands's  result~\cite{Wands:1998yp}, can be found below\footnote{%
Wands's result can be re-casted into  a simplified form,
$\mathcal{P}_{\delta\phi}\, \sim\, H^2(-k\eta)^{3-2|\beta|}$
with
$\mathcal{P}_{\delta\phi}$, $H$, $k$ and $\eta$
denoting a given power spectrum, the Hubble parameter, wave vector and conformal time respectively.
And $\beta=\pm\frac{3}{2}$ corresponds to an expanding de-Sitter background and a matter-dominated contraction, respectively.
Naively for both cases
$\beta=\pm\frac{3}{2}$, the $\mathcal{P}_{\delta\phi}$ are $k$-independent.
However, $H$ should be measured at the moment at which a given perturbation mode exits its effective horizon and hence picking up a dependence in time. For an expanding de-Sitter background, the Hubble parameter, $H$, is the same for each mode as it exits its effective horizon; while in a matter-dominated contraction, the value of the Hubble parameter varies with time and therefore each perturbation mode takes on a different value of $H_*(k)$ as the mode exits its effective horizon. 
In other words,  a $k$-dependence enters the final expression of power spectrum through $H_*(k)$ in Wands' dual solution  in a matter-dominated contraction.}.

We pause here to clarify our terminologies.
A  power spectrum of density perturbations, $\mathcal{P_\chi}$,  is conventionally defined as follows,
\begin{equation} \label{eq-spectrum}
\mathcal{P_\chi}\, =\, \frac{k^3 \, |\chi|^2}{2\pi}~.
\end{equation}
Since we are interested in the deviation from scale invariance  we will suppress the three powers of $k$ leading to  scale invariance and concentrate on any deviations from it. 
For a gaussian spectrum we will simply write:
\begin{equation}
P_\chi\equiv |\chi|^2~.
\end{equation}

Scale-invariance ($k$-independence) of a power spectrum of $\chi$ then implies, 
\begin{equation} \label{eq-minew}
\mathcal{P_\chi}\sim k^0 \,\eta^{2W(\nu,m)} \quad 
\leftrightarrow \quad P_\chi\sim k^{-3}\,\eta^{2W(\nu,m)}
\end{equation}
where $2W(\nu,m)\ne0$, in general.
Magnitude-invariance ($\eta$-independence) of a power spectrum of $\chi$, likewise, implies
\begin{equation} 
\label{eq-sinew}
\mathcal{P_\chi}\sim k^{2L(\nu,m)+3}\, \eta^0\quad 
\leftrightarrow \quad P_\chi\sim k^{2L(\nu,m)}\,\eta^0~,
\end{equation}
where $2L(\nu,m) \ne\,0 $, in general.
The  two equations above also serve to  define  $2L(\nu,m)$ and $2W(\nu,m)$.  The detailed derivations  of these two indices  from the dynamic equations are given below in Section~\ref{sec:dynamics}.

A power spectrum can be scale independent in a physical sense  if and only if it is both time independent and scale independent: 
\begin{equation} \label{eq-sminew}
\mathcal{P_\chi}\sim k^0\, \eta^0 \quad 
\leftrightarrow\quad  
P_\chi\sim k^{-3}\, \eta^0~,
\end{equation}

We conform to the conventional usage of   ``scale-invariance''  in denoting a power spectrum's  $k$-independence, and use ``magnitude-invariance" to stress the lack of time evolution of the magnitudes. 
We would like to stress that  only when a solution that is both ``magnitude-invariant" and ``scale-invariant" in the conventional sense can  truly be a scale-independent power spectrum. 
In the following, we will see that the power spectrum generated in the single scalar field slow-roll inflation model is a genuine  scale-independent power spectrum. Unfortunately, in the Wands' case, the power spectrum is only scale-invariant but not magnitude-invariant.

Followed from the above discussion it is obvious that we should check the time evolution of each individual mode in a power spectrum to see if its magnitude  varies with time.
To this end we extend the usual blue/red-shift term, $3H\, \dot\chi$ 
in the dynamical equation for the  gauge invariant perturbations $\chi$ to a more  general  form, $mH\dot\chi$. 
This generalized  blue/red-shift is found in many a cosmological model, {\it e.g.} the bounce universe models~\footnote{See~\cite{bouncereview} for a review.},  the multi-field models with non-standard kinetic term and non-trivial potential~\footnote{A general formalism for multi-scalar-fields models with second-order actions of cosmological perturbations  can be found in~\cite{multifield}.} etc.. 
Adopting a general blue/red-shift term enables us to unveil the $(\nu, m)$ parameter space of power spectra, where $\nu$ is the power law index of the scale factor of the background defined as $a=\eta^{\nu}$ with $\eta$ being the conformal time. 
Each point in the $(\nu,m)$ space hence corresponds to one cosmological model with  the perturbation modes  evolving  according to, $mH\, \dot\chi$, and  the  background according to, $a=\eta^\nu$.

\fig{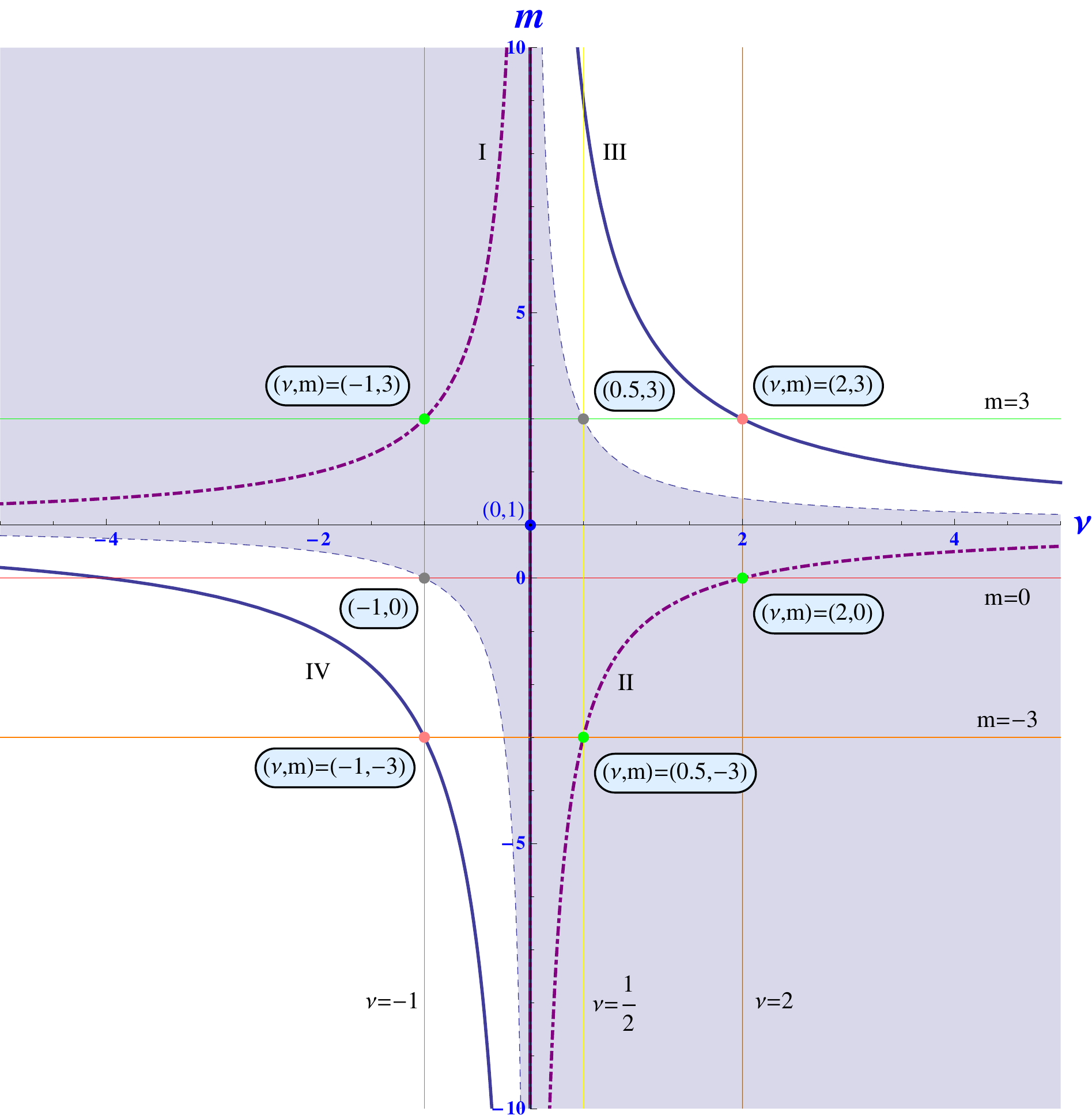}{%
The $(\nu,m)$ parameter space for classifying scale invariant and magnitude invariant solutions of the power spectra of density perturbations. $\nu$ is the power law index (horizontal axis) and $m$ is  the red/blue-shift index (vertical axis).
The shaded region includes all magnitude-invariant solutions satisfying $(m-1)\,\nu-1 < 0$  whose boundaries are defined by $(m-1)\,\nu-1= 0$ and drawn with thin dash lines. 
The purple dot-dash lines obeying $(m-1)\,\nu =-2$  give  physically scale-invariant solutions. Another set of scale invariance solutions given by  
$(m-1)\,\nu =4 $ (violet solid lines) have Fourier modes varying with time and therefore are not truly scale-invariant in a physical sense.%
}{0.8}{h!t}

As explained in detail in Section~\ref{SIMISEC}  power spectra can be classified by a two dimensional  parameter space shown below in~\refig{vmmap126.pdf}  with $\nu$ being the power law index (horizontal axis) and $m$ the red/blue-shift index (vertical axis). It is then easy to notice the following four salient features: 
\begin{enumerate}
\vspace{-0.3cm}
\item[1.]
We find, to our surprise, that the magnitude-invariant  solutions  occupy a two dimensional region in the $(\nu, m)$ parameter space, rather than some one-dimensional curves as one would naively expect.
Dynamically, $\nu$ and $m$ serve as two independent and free parameters for the equation of motion for the  perturbations obeying~\refeq{chit} and~\refeq{ac}, respectively. 
Therefore, the $\eta$-dependence and $k$-dependence of the spectrum of $\chi$  are completely determined by a point in the   $(\nu, \, m)$ space as given by  $\mathcal{P}_\chi \sim k^{2L(\nu,m)+3}\,\eta^{2W(\nu,m)}$. 
The power spectra satisfy the relations $2L(\nu,m)+3=0$ (~\refeq{l}) and $2W(\nu,m)=0$ (\refeq{w}) are, respectively,  scale-invariant and magnitude-invariant.  

Generically, one would  expect each of  these two relations lead to one, or more,  one dimensional curve in the $(\nu,m)$ parameter space, and these curves have one or several intersection  points which can give rise  to some scale-invariant  and          magnitude-invariant solutions.  
From this point of view, the magnitude-invariant and scale-invariant solutions should distribute sparsely  in the $(\nu,m)$  space. 
However, thanks to a remarkable duality of the equation of motion for $\chi$, the magnitude-invariant relation, $2W(\nu,m)=0$, is an inequality of $\nu$ and $m$ rather than an equation of constraint (\refeq{we} and \refeq{mi}). 
Mathematically, an inequality often allows a larger set of solutions. 
In this case, the solution space~\refeq{mi} of magnitude-invariant power spectra  occupies a two dimensional region. 
In~\refig{vmmap126.pdf}, the magnitude-invariant region is the shaded area, and its boundaries, $(m-1)\,\nu-1= 0$, are depicted by a pair of  thin dash lines. 

Theoretically, this novel property implies that the probability of getting a magnitude-invariant solution by choosing a random value of $(\nu,m)$ is nearly 
$\frac{1}{2}$~--in contrast to a measure zero set as previously expected. 
Moreover, given the two curves of the scale-invariant solutions lie in the two dimensional magnitude-invariant region, the number of scale-invariant and magnitude-invariant solutions are infinite rather than finite.
A substantial portion of the cosmological models can henceforth produce a 
magnitude-invariant power spectra. And with careful choice of $\nu$ and $m$ 
producing a power spectrum with both scale-invariance  and magnitude-invariance 
is no longer  a monopoly of the  slow-roll inflation scenario!

\item[2.]
There are two groups of scale-invariant solutions, one of which is magnitude-invariant but  the other  changes  with time. 
According to the scale-invariant relation $2L(\nu,m)+3=0$, we have two groups of scale-invariant solutions, 
$(m-1)\,\nu =-2$ (purple dot-dash lines) and $(m-1)\,\nu =4 $ (violet solid lines). 
The  scale-invariant solutions, $(m-1)\,\nu =-2$, are also magnitude-invariant. 
Only solutions, which are both scale-invariant and magnitude invariant, are truly independent of $k$. They therefore lie inside of the magnitude-invariant region  
(shaded purple), the boundary (the dash-dot lines)  of which  obeys $(m-1)\,\nu-1= 0$  in~\refig{vmmap126.pdf}. 
The other group of scale-invariant solutions, $(m-1)\,\nu =4 $ (violet solid lines), are magnitude-non-invariant. Cosmologically, the time-dependence of their power spectrum render them implicitly $k$-dependent, so that they are not truly scale-invariant. 
In~\refig{vmmap126.pdf} they lie on the pair of solid blue lines but outside of the magnitude-invariant region. 

\item[3.]
A novel duality between the scale-invariant solutions and the boundaries of the magnitude-invariant solutions can be detected   from a closer inspection 
of~\refig{vmmap126.pdf}. 
We notice that the fourth vertex of a rectangle, the  three  other vertices of which lie on the scale-invariant solutions curves, is located on the boundary of 
magnitude-invariant region. 
For instance, the fourth vertex of the rectangle with other three vertexes being $(-1,3)$, $(2,3)$ and $(2,0)$  (marked pale blue), is located on the $\nu<0$ branch of the boundary $(m-1)\,\nu-1 = 0$.  
This in turn  implies the existence of a duality, in the same light of that found by Wands~\cite{Wands:1998yp},  between the scale-invariant solutions and the boundary of the magnitude-invariant region.

Generically,  starting from a scale-invariant and magnitude-invariant solution, under two consecutive duality transformations defined in Section~\ref{sec:duality}, we can obtain an accompanying scale-invariant and magnitude-invariant solution. This scale-invariant and magnitude-invariant solution and its accompanying solution, therefore, determine a solution of the boundary of the magnitude-invariant region uniquely. Moreover, going through all scale-invariant and magnitude-invariant solutions, we are able to determine the whole boundary of the magnitude-invariant region. Though the physical origin of such a duality is yet to be explored, we show the existence of  this duality with explicit calculation in the Section~\ref{sec:duality}.

\item[4.] %
Several distinguished solutions, corresponding  to several important cosmological models, are included in the $(\nu, m)$ parameter space.
Studying them case by case, we find that the three points, $(-1,3)$, $(2,3)$ and $(2,0)$,  correspond to, in order, the well-known single field slow-roll inflation model, Wands' matter-dominated contraction model and an Ekpyrotic model with negligible  gravitational back-reaction.  
They are, in the same order, scale-invariant and magnitude-invariant, scale-invariant and magnitude-non-invariant, and magnitude-invariant and scale-non-invariant, and are consistent with   previously known results.
We can easily predict the existence of  a matter-dominated contraction solution, $(0,2)$, with both scale-invariance as well as magnitude-invariance. This solution is nothing but   the power spectrum of the tachyon field in the coupled tachyon-scalar  bounce universe model~\cite{Li:2011nj} proposed earlier. 
 
%
\end{enumerate}

This paper is organized as follows. 
In Section~\ref{sec:dynamics}, we compute the expression of power spectrum of density perturbations in terms of $k$ and $\eta$ with general blue/red-shift term $mH\dot{\chi}$. 
In Section~\ref{SIMISEC}, we derive the magnitude-invariant and scale-invariant solutions, and discuss the general duality of $(\nu, m)$ parameter space. 
In Section~\ref{sec:duality}, we give a proof to the duality between scale-invariant solution and the boundaries of magnitude-invariant solutions. 
In Section~\ref{sec:CaseStudy}, we apply our general results to a few examples of interest, namely, the de-Sitter background, the Ekpryotic  (slowly contracting/expanding)  background, and finally a matter-dominated  contracting  background. 
And, last but not  least,  we summarize our findings and discuss their implications to cosmological model building in  Section~\ref{sec:disc}.

\section{The dynamical equations for density perturbations}
\label{sec:dynamics}

For our purpose $\chi$ is taken to be a gauge invariant variable related to the linear density perturbations of an underlying cosmological model. 
The equation of motion for  each  Fourier mode, $\chi_k$, of, $\chi$, then takes the following form,
\begin{equation} 
\label{eq-chit}
\ddot{\chi}_k\, +\, m H\dot{\chi}_k\, +\, \frac{k^2}{a^2}\, \chi_k\, =\, 0~.
\end{equation}

The cosmological background is, meanwhile, evolving by a power-law
\begin{equation} \label{eq-ac}
a\propto \eta^\nu~,  
\end{equation}
where $a$ being the scale factor and $\eta$ the conformal time.
We will henceforth call, $\nu$, the ``power-law index'' of the underlying cosmological background. 
It is then obvious that $\eta\rightarrow 0$  corresponds to era in which perturbations could exit their ``effective'' horizon  while $\eta\rightarrow\infty$ corresponds to perturbations re-entering the ``effective'' horizon. 
In this epoch with $\eta\le 0$  an expanding phase is described by $\nu<0$ while the contracting phase  is described by $\nu>0$.

Therefore the solutions of $\chi$ for a given cosmological model are expected to locate on the two dimensional parameter space$(\nu,m)$. We can also interpret the $(\nu, m)$ in terms of the ``effective'' equation of state of $\chi$ and that of the cosmological background, $(\omega_b, \omega_\chi)$, which are related in a flat FLRW background by
\begin{equation}
\nu=\frac{2}{3(1+\omega_b)-2}
\end{equation}
and
\begin{equation}
m=3(1+\omega_\chi)~.
\end{equation}

The physical implication of our starting point is that the
``effective" Equation of State of the perturbations as well as those
of the cosmological background fields would be different, for
instance, in a multi-field cosmological model, in models with
nonlinear kinetic
terms~\cite{Garriga:1999vw}\cite{ArmendarizPicon:2000ah}\cite{Khoury:2011ii},
or in models with
quintessence~\cite{Zlatev:1998tr}\cite{Chiba:1999ka}\cite{Malquarti:2002iu}
and so on. Of course the best known example is still the slow-roll
inflation model which we will analyze in detail as a concrete
application to our proposal. We will use it as a reference point to
compare other models whenever possible.

With the following change of variables (we use $u$ instead of $u_k$ as there is no ambiguity in meaning.):
\begin{equation} 
\label{eq-xau}
\chi_k\equiv a^{-\frac{1}{2}(m-1)}u 
\end{equation}
allowing us to  simplify \refeq{chit} 
\begin{equation} \label{eq-eomu}
u^{\prime\prime}+(k^2-\frac{2\gamma}{\eta^2})u=0 
\end{equation}
where
\begin{equation} \label{eq-gamma}
\gamma\equiv \frac{1}{2}\left[\frac{(m-3)(m-1)}{4}\nu^2 - \frac{1-m}{2}\nu(\nu-1)\right] 
\end{equation}
For non-zero $\gamma$, when $\eta\rightarrow 0$, $u$ of different wave vectors
 will gradually get out of  their  ``effective horizon'' defined by
\begin{equation}
\left|-\frac{2\gamma}{\eta^2}\right| =  k^2~.
\end{equation} 
The general solution for $u$ then  
takes the familiar form
\begin{equation} \label{eq-usoln}
u=C_j(\gamma)\eta^{\frac{1}{2}}J_n\left(k\eta\right)+C_y(\gamma)\eta^{\frac{1}{2}}Y_n\left(k\eta\right)
\end{equation}
where $J_n$ and $Y_n$ are the $n$-th order Bessel functions, of first and second kind, respectively, with
\begin{equation}
n\equiv \frac{1}{2}\sqrt{1+8\gamma}~,
\end{equation}
with $C_j$ and $C_y$ being the coefficients that could depend on $\gamma$ but not on $k\eta$.

We observe that~\refeq{gamma} yields the condition:
\begin{equation} \label{eq-egamma}
1+8\gamma=\left[(m-1)\,\nu-1\right]^2\ge0~,
\end{equation}
and thus in the limit $k\eta\rightarrow 0$ the leading order contribution to $u$ becomes
\begin{equation} \label{eq-ulo}
u\sim C(\gamma) \eta^{\frac{1}{2}}(k\eta)^{-\frac{1}{2}\sqrt{1+8\gamma}}~.
\end{equation}

Substituting \refeq{ac} and \refeq{xau} into \refeq{ulo},  we have
\begin{eqnarray}  
\label{eq-klo}
\chi_k  
&\sim & C(\gamma) \eta^{-\frac{1}{2}(m-1)\,\nu}\eta^{\frac{1}{2}}(k\eta)^{-\frac{1}{2}\sqrt{1+8\gamma}}  \\ \nonumber
&\sim  &  C(\gamma) \eta^{-\frac{1}{2}\left[(m-1)\,\nu-\left(1-\sqrt{1+8\gamma}\right)\right]}k^{-\frac{1}{2}\sqrt{1+8\gamma}}\\ \nonumber
&\sim  &  C(\gamma) \eta^{W(\nu, m)}k^{L(\nu,m)}
\end{eqnarray}
where we have introduced the magnitude-index, $W(\nu,m)$, 
\begin{equation} 
\label{eq-w}
W(\nu,m)\equiv-\frac{1}{2}\left[(m-1)\,\nu-\left(1-\sqrt{1+8\gamma}\right)\right]~, 
\end{equation}
and the scale-index,  $L(\nu,m)$, 
\begin{equation} 
\label{eq-l}
L(\nu,m)\equiv -\frac{1}{2}\sqrt{1+8\gamma}.
\end{equation}

\section{The scale-invariance and magnitude-invariance conditions for power spectra of cosmological perturbations}
\label{SIMISEC}

In this section, we discuss the  conditions under which a power spectrum  of  cosmological perturbations $\chi$ could be scale invariant and/or magnitude invariant.
A power spectrum $\mathcal{P_\chi}$ is conventionally defined as following,
\begin{equation}
\mathcal{P_\chi}\equiv \frac{k^3 |\chi|^2}{2\pi}
\end{equation}
And in this paper, we sometimes use
\begin{equation}
P_\chi\equiv|\chi|^2
\end{equation}
as well to concentrate on the  deviation from usual scale invariant spectrum.
A power spectrum written in terms of the underlying variables, $k$ and $\eta$, becomes:
\begin{equation} \label{eq-Pketa}
\mathcal{P_\chi}\sim k^{2L(\nu,m)+3} \, \eta^{2W(\nu,m)}, \quad i.e.
\quad P_\chi\sim k^{2L(\nu,m)}\eta^{2W(\nu,m)},
\end{equation}
Consequently  the scale-invariance ($k$-independence) of a power spectrum of $\chi$ implies
\begin{equation} \label{eq-minew}
\mathcal{P_\chi}\sim k^0\eta^{2W(\nu,m)}, \quad i.e.\quad P_\chi\sim k^{-3}\eta^{2W(\nu,m)},
\end{equation}
and the magnitude-invariance ($\eta$-independence)  implies
\begin{equation} \label{eq-sinew}
\mathcal{P_\chi}\sim k^{2L(\nu,m)+3}\eta^0, \quad i.e.\quad P_\chi\sim k^{2L(\nu,m)}\eta^0.
\end{equation}
Finally  a  physically scale-invariant (both $k$ and $\eta$ independent) power spectrum is the one satisfies
\begin{equation} \label{eq-sminew}
\mathcal{P_\chi}\sim k^0\eta^0, \quad i.e.\quad P_\chi\sim k^{-3}\eta^0~.
\end{equation}

\subsection{Magnitude Invariance condition}

The magnitude-index defined by~\refeq{w} can be expressed as
\begin{equation} \label{eq-we}
W(\nu,m)=-\frac{1}{2}\left\{(m-1)\,\nu-1 \, +\left|(m-1)\,\nu-1\right|\right\} 
\end{equation}
by substituting \refeq{gamma} into \refeq{w}. Magnitude-invariance then implies: 
\begin{equation} \label{eq-wz}
W(\nu,m)=0~.
\end{equation}
Therefore the spectra of cosmological perturbations, $\chi$, 
whose values of $(\nu, m)$ obey 
\begin{equation} 
\label{eq-mi}
(m-1)\,\nu-1\le 0~
\end{equation}
are  magnitude-invariant  solutions. 
These magnitude-invariant power spectra span a two dimensional magnitude-invariance region, shaded violet in~\refig{vmmap126.pdf}, 
and the boundary of magnitude-invariance region is given by 
$\displaystyle{m=\frac{1}{\nu}+1}$, 
as delineated by a pair of grey dashed lines. 
In term of the equations of state, $\omega_b$ and $\omega_\chi$, the boundaries are expressed as
\begin{equation}
\omega_\chi=\frac{1}{2}(\omega_b-1)~.
\end{equation}

\subsection{Scale Invariance condition}

Likewise, from~\refeq{sinew}, we  obtain the scale-invariance condition,
\begin{equation}
2L(\nu,m)+ 3 = 0 \Longrightarrow \gamma=1,
\end{equation}
and the solutions are
\begin{equation}   
\label{eq-simi} 
m = \left\{  
  \begin{array} {l}
 {\displaystyle\frac{4}{\nu}+1}   \\ 
 \\ 
  {\displaystyle-\frac{2}{\nu}+1}    \\
\end{array}     
\right.
\end{equation}
both combinations of $\nu$ and $m$ lead to scale-invariant spectra of $\chi$. 

The  scale-invariant solutions from  $\displaystyle{m=-\frac{2}{\nu} +1 }$ 
branches--indicated by purple dot-dash lines in~\refig{vmmap126.pdf}--also satisfy the magnitude invariance condition
\begin{equation}
(m-1)\,\nu-1\le 0~.
\end{equation}
Therefore we conclude that  $\displaystyle{m=-\frac{2}{\nu}+1}$ branch gives  physically scale-invariant solutions. But the scale-invariant solutions on the  $\displaystyle{m=\frac{4}{\nu}+1}$ branches--indicated by solid violet lines in~\refig{vmmap126.pdf}\,--are not truly scale-invariant solutions as the magnitudes of their Fourier modes  change with conformal time, $\eta$. This is corroborated by the fact that the pair of lines lie outside the magnitude invariant region.

Scale-invariant solutions~\refeq{simi} can also be expressed in terms of equations of state of the background, $\omega_b$, and of  the field,  $\omega_\chi$ as follows:
\begin{equation}
\omega_\chi=\omega_b
\end{equation}
and
\begin{equation}
\omega_\chi=-(1+\omega_b)~.
\end{equation}

\section{Duality between the scale-invariant solutions and the boundary of the magnitude-invariance region}
\label{sec:duality}

In this section, we will unveil a novel duality between scale-invariant solutions and the boundary of the magnitude-invariance region.

For each $m$, there are two scale-invariant solutions, $(\nu, m)$ and $(\hat\nu, m)$, satisfying the two scale-invariance conditions~\refeq{simi}, 
\begin{equation}
m=\frac{4}{\nu}+1, \quad {\mathrm{or}}, \quad m=-\frac{2}{\hat{\nu}}+1~.
\end{equation}
Two solutions having the same value of $m$  implies the dynamics equation of the perturbations $\chi$ for these two solutions are the same. 
Therefore  we call the duality between these two scale-invariant solutions as ``Iso-Perturbation Scale-Invariance (IPSI)" duality,  and a transformation between them
``Iso-Perturbation Scale-Invariance" transformation.

Similarly, for each  $\nu$,  there are two scale-invariant solutions,  
$(\nu, m)$ and $(\nu, \hat{m})$, satisfying the either one of  scale-invariance conditions~\refeq{simi}
\begin{equation}
m=\frac{4}{\nu}+1, \quad {\mathrm{or}}, \quad \hat{m}=-\frac{2}{\nu}+1~.
\end{equation}
The same value of $\nu$ implies that the perturbations $\chi$ for these two solutions evolve in the same background, we call the duality between these two scale-invariant solutions as ``Iso-Background Scale-Invariance (IBSI)"  and the transformation between them as ``Iso-Background Scale-Invariance''  transformation.

With the IPSI and IBSI dualities at hand, we are able  to show the existence of a subtle duality between the scale-invariant solutions and the boundary of the magnitude-invariance region.  We summarize our findings in the following theorem.

\begin{theorem}
The boundary of the magnitude-invariance region can be determined uniquely by 
from  scalar-invariant solutions by two Scale-Invariance transformations.
\end{theorem}

\textit{Proof}: Without loss of generality, we start from the point $(\nu_1, m_1)$ on the $\nu<0$ quadrant of the scalar-invariant and magnitude-invariant branch of $m=-\frac{2}{\nu}+1$, say $(-1,3)$ for concreteness.
We shall  label the IPSI duality transformation as $D_{IPSI}$ and IBSI duality transformation as $D_{IPSI}$. 
Since $D_{IPSI}$ and $D_{IBSI}$ may not be commutative, we act both $D_{IBSI}D_{IPSI}$ and $D_{IPSI}D_{IBSI}$ on the $(\nu_1, m_1)$ respectively at following.
In the $D_{IPSI}D_{IBSI}$ case, acting $D_{IBSI}$ on $(\nu_1, m_1)$ gives,
\begin{equation}
D_{IBSI}(\nu_1, m_1) =(\nu_1, m_2),
\end{equation}
taking $(-1,3)$ to $(-1,-3)$, landing on another scale-invariant solution on the 
\begin{equation}
m_2=\frac{4}{\nu_1} +1~
\end{equation}
branch, which violates the magnitude-invariance condition, 
$(m-1)\,\nu-1\le 0$.
Continue acting with $D_{IPSI}$ on $(\nu_1, m_2)$,
\begin{equation}
D_{IPSI}(\nu_1, m_2)=(\nu_2, m_2)
\end{equation}
maps it to another  the scale-invariant solution $(0.5, -3)$ on the $\nu > 0$ branch of the  
$$\displaystyle{m_2 = -\frac{2}{\nu_2} + 1 }$$ 
curve which lies properly inside the shaded region of magnitude invariant solutions. 
Therefore $(\nu_2, m_2)$ satisfies both the scale-invariance as well as the  magnitude-invariance conditions.

Both $ (\nu_1, m_1)$  and $ (\nu_2, m_2)$ (These points are $(-1,3)$ and $(0.5,-3)$ in the example we give.)  are scale-invariant and magnitude-invariant in this case. And more intriguingly  the point $(\nu_2, m_1)$  satisfies the boundary condition of the magnitude-invariance region:
\begin{equation}
m_1=\frac{1}{\nu_2}+1~
\end{equation}
 hence  lies  on the grey line of the shaded region.
Therefore  we can map out  the boundary of magnitude-invariant solutions 
in  the $(\nu>0, m>0)$  quadrant by acting with consecutive Iso-Background and 
Iso-Perturbation duality transformations, $D_{IPSI}D_{IBSI}$,  on the solutions lying on the purple dash dot line--both  scalar-invariant and magnitude-invariant--in  
the $(\nu<0, m>0)$ quadrant.
The above observation also hints at a duality between a scale-invariant 
solution $(-1,-3)$ to a magnitude invariant solution $( \, 0.5, 3\, )$ 
which warrants further study for model building purposes.

We now turn our attention to the  $D_{IBSI}D_{IPSI}$ case, i.e. one Iso-Perturbation  transformation followed by one Iso-Background  transformation.
Acting $D_{IPSI}$ on $(\nu_1, m_1)$
\begin{equation}
D_{IPSI} (\nu_1, m_1) = (\tilde\nu_2, m_1),
\end{equation}
we obtain the $(\tilde{\nu}_2, m_1)$ (mapping $(-1,3)$ to $(2,3)$)
 which  is also a scale-invariant solution lying on the $\nu > 0$ branch of 
\begin{equation}
m_1=\frac{4}{\tilde\nu_2} + 1~,
\end{equation}
but violates  magnitude-invariance condition, $(m-1)\,\nu-1\le 0$, and hence lying outside of the purple shaded region.

Then we act $D_{IBSI}$ on $ (\tilde\nu_2, m_1)$,
\begin{equation}
D_{IBSI} (\tilde{\nu}_2, m_1)=(\tilde{\nu}_2, \tilde{m}_2)
\end{equation}
to $ (\tilde{\nu}_2, \tilde{m}_2)$  (mapping $(2,3)$ to $(2,0)$), which satisfies both the scale-invariance and magnitude-invariance conditions,
\begin{equation}
\tilde{m}_2=-\frac{2}{\tilde{\nu}_2}+1 \quad {\mathrm{and}} \quad (m-1)\,\nu-1\le 0~,
\end{equation}
respectively.
Both $(\nu_1, m_1)$ and $ (\tilde{\nu}_2, \tilde{m}_2)$ 
(in our example   $(-1,3)$ and  $(2,0)$ ) 
are scale-invariant and magnitude-invariant. 
The  point $(\nu_1, \tilde{m}_2)$ 
lies on  the boundary  of magnitude-invariant solutions,
\begin{equation}
\tilde{m}_2=\frac{1}{\nu_1}+1~.
\end{equation}
Going along the dot-dash line in the  $(\nu<0, m>0)$ quadrant, acting with 
$D_{IPSI}D_{IBSI}$, we can map out the boundary of magnitude-invariant solutions 
in the $(\nu<0, m<0)$ quadrant of the $(\nu, m)$ parameter space. 

Clearly seen from the above analysis the IBSI duality  and IPSI duality transformations do not commute. $D_{IPSI}D_{IBSI}$ and $D_{IBSI}D_{IPSI}$  acting on the scalar-invariant solutions will map out for us, respectively, the upper  
and lower 
boundaries of magnitude-invariant solutions~\refig{vmmap126.pdf}.

\section{General Duality}
\label{sec:general}

In this section, we are studying the general duality, which leave $k$-dependence and/or $\eta$-dependence of power spectrum unchanged, in the $(\nu, m)$ parameter space.
The expressions of the scale-index, $L(\nu,m)$~\refeq{sinew}, and the 
magnitude-index, $W(\nu,m)$~\refeq{we}, in terms of $\nu$ and $m$ are
\begin{equation} \label{eq-le}
L(\nu,m)=-\frac{1}{2}\left|(m-1)\,\nu-1\right|,
\end{equation}
and
\begin{equation}
W(\nu,m)=-\frac{1}{2}\left\{\left[(m-1)\,\nu-1\right]+\left|(m-1)\,\nu-1\right|\right\}
\end{equation}
Then a power spectrum with $(\hat{\nu},\hat{m})$, which satisfies
\begin{equation}
\label{eq-gkd}
\left|(m-1)\,\nu-1\right|=\left|(\hat{m}-1)\hat{\nu}-1\right|,
\end{equation}
will have same $k$-dependence as  a power spectrum with $(\nu, m)$.
According to~\refeq{we}, a power spectrum with $(\tilde{\nu},\tilde{m})$,  satisfying
\begin{equation} \label{eq-gwd}
\left[(m-1)\,\nu-1\right]+\left|(m-1)\,\nu-1\right|=\left[(\tilde{m}-1)\tilde{\nu}-1\right]+\left|(\tilde{m}-1)\tilde{\nu}-1\right|,
\end{equation}
has same $\eta$-dependence with a power spectrum with $(\nu, m)$.

Furthermore if  $(m-1)\nu-1$ and $(\hat{m}-1)\hat{\nu}-1$  are either both positive or both negative  simultaneously, then they can also satisfy the last requirement~\refeq{gwd}. When this is the case,  a power spectrum with $(\hat{m}-1)\hat{\nu}-1$ will have the same $k$-dependence as well as  $\eta$-dependence with a power spectrum with $(\nu, m)$. Note, however, that when they are both positive they can neither be scale-invariant nor magnitude-invariance.   We are now ready to proceed to a detailed case study.

\section{Case Study}
\label{sec:CaseStudy}
In this section, we will relate our results to the several important cosmological models--putting into context to help illustrate the power of dualities.
This also helps in the search of   new cosmological models which can  produce scale-invariant solutions. 

\subsection{de-Sitter Inflation Background}
In de-Sitter cosmological background, we have $\nu=-1$. 
According to the scale invariance condition, \refeq{simi} 
there are two power spectra satisfying the condition yielding two scale-invariant solutions $(\nu,m)=(-1,3)$ and $(\nu,m)=(-1,-3)$. 
Furthermore, the $(-1,3)$ case  is also magnitude-invariant following from   the magnitude-invariance condition~\refeq{mi}. $P_\chi(1,-3)\propto \eta^0$.  
On the other hand  the $(-1,-3)$ case is not  magnitude-non-invariant as  $P_\chi(-1,-3)\propto \eta^{-3}$. 

An alert reader would have recognized  $(-1,3)$  as  the celebrated  single field slow-roll inflation model. The evolution of its linear perturbations of scalar (the inflaton), $\delta \phi$ (in spatially-flat slicing), and the  tensor mode $h_{+/\times}$  
are governed by~\refeq{chit} and~\refeq{ac} 
with $(\nu,m)\simeq(-1,3)$~\cite{Dodelson:2003ft}. 
Therefore, according to above analysis, the power spectrum of $h_{+/\times}$ and $\delta \phi$ should be nearly scale-invariant and magnitude-invariant. The tensor mode of perturbation $h_{+/\times}$ is gauge-invariant, but $\delta\phi$ is not gauge-invariant. Fortunately, $\delta\phi$ relates to the gauge-invariant quantity $\zeta$, which is related to  the scalar  perturbations  by a factor  $\frac{aH}{\dot{\phi}}$,
\begin{equation}
\zeta=-\frac{aH}{\dot{\phi}}\delta\phi~.
\end{equation}
So the power spectrum of $\zeta$ is also nearly scale-invariant: the conversion factor gives only a small contribution to the running of  spectral index.

In the single scalar field slow-roll model, we have $(\bar{\nu},\bar{m})=(-1,3)$ and $\delta m=0$. Then the primordial spectrum of tensor mode, $h_{+/\times}$, becomes
\begin{equation}
P_h\, \sim\, k^{-\left|\left[(\bar{m}-1)\bar{\nu}\right]+\bar{\nu}\delta m+(\bar{m}-1)\delta\nu\right|}\,\sim\, k^{-3+2\delta\nu}.
\end{equation}
We obtain the primordial spectral index $n_T$ of the tensor mode
\begin{equation}
n_T\equiv\frac{d\ln{(k^3P_h)}}{d\ln{k}}=2\delta\nu \label{eq-ntn}
\end{equation}

The slow-roll parameter, $\epsilon$, is defined as,
\begin{equation}
\epsilon\equiv\frac{d}{ad\eta}\left(\frac{1}{H}\right)=\frac{\nu+1}{\nu}\simeq-\delta\nu \label{eq-en}
\end{equation}
Substituting~\refeq{en} into~\refeq{ntn}, we re-derive  the well-known relation between the $n_T$ and $\epsilon$ for the slow-roll inflation model
\begin{equation} \label{eq-nte}
n_T=-2\epsilon~.
\end{equation}
In above analysis we do not need to know at which exact time the $k$ mode exit the horizon since such information have been encoded into~\refeq{klo} through~\refeq{gamma}.

For the single field slow-roll inflation model, the factor $-\frac{aH}{\dot{\phi}}$ is related to the slow parameter $\epsilon$,
\begin{equation}
\left(-\frac{aH}{\dot{\phi}}\right)^2\propto\epsilon^{-1}=(-\delta\nu)^{-1}.
\end{equation}
Therefore, we obtain the primordial spectra index of scalar mode $n_s$,
\begin{eqnarray}
\nonumber n_s-1\equiv \frac{d\ln{(k^3P_\zeta)}}{d\ln{k}}\qquad\qquad\qquad\qquad\qquad\quad\\=\frac{d\ln{(k^3P_{\delta\phi})}}{d\ln{k}}-\frac{d\ln{(-\delta\nu)}}{d\ln{k}}=2\delta\nu-\frac{\delta\nu^\prime}{\delta\nu}\eta \label{eq-nsn}
\end{eqnarray}
With the~\refeq{ntn} and~\refeq{nsn}, the non-zero values of primordial spectral indices, $n_T$ and $n_s-1$, reflect the small departure of the  cosmological background of the single field slow-roll inflation  from a perfectly  de-Sitter cosmos.

We remark here  that $\delta m=0$ for the single scalar field slow roll inflation since its kinetic term is simply  $X=\frac{1}{2}\dot{\phi}^2$. 
For a more general kinetic term, $P(X)$, the $\delta m$ could not be neglected. In that case, the $\delta m$ will appear in the expression of $n_T$ and $n_s$.  In other words  precise measurements  $n_T$ and $n_s$ can be used to resolve the present degeneracy of cosmological models.  

\subsection{Slowly Expanding/Contracting Backgrounds}

An Ekpyrotic/cyclic scenario requires a slowly expanding or contracting 
background~\cite{Khoury:2001zk} which can be characterized by $\left|\nu\right|\ll1$. 
We will now study  the power spectrum of cosmological perturbations in such a background.

The potential of the single scalar field, in this scenario, takes the  following form:
\begin{equation}
V(\phi)=-V_0 exp(-\sqrt{\frac{2}{p}}\frac{\phi}{M})~,  
\end{equation}
with $M\equiv\sqrt{\frac{1}{8\pi\,G}}$ as usual; because $p \ll 1 $, the potential very steep. 
The equation of motion governing the cosmological perturbations~\cite{Creminelli:2004jg} is,
\begin{equation}
\ddot{\xi}+3H\dot{\xi}+\frac{k^2}{a^2}\xi=0
\end{equation}
where $\xi\equiv-\sqrt{\frac{p}{2}}\frac{\delta \phi}{M}$; 
meanwhile the cosmological background obeys
\begin{equation}
a\propto t^p ~~~\rightarrow~~~ a \propto \eta^\frac{p}{1-p}~.
\end{equation}
One then gets  $(\nu,m)=(\frac{p}{1-p}, 3)$, 
giving  a power spectrum of $\xi$,
\begin{equation}
P_\xi\propto \eta^0\, k^{-\frac{1-3p}{1-p}}~.
\end{equation}
The spectrum $P_\xi$ is time-independent (i.e. magnitude-invariant) as  expected.  For $0<p\ll1$, 
\begin{equation}
P_\xi \sim k^{-1 + \frac{2p}{1-p} } \sim k^{-1}~.
\end{equation}
Therefore, $P_xi$ is not scale-invariant, but it is red-tilted in comparison with a scale-invariant spectrum  
$P \sim k^{-3}$. These results are in accordance  with  the previously  analysis~\cite{Creminelli:2004jg}.
With the scale invariance conditions~\refeq{simi}  the scale-invariant power spectrum cannot be generated in a slowly expanding/contracting background unless 
$\left| m \right|$ is very large ($m \sim -(2p)^{-1}$), which is made trivially obvious in our language.

\subsection{Matter-dominated contracting background%
}
In a matter-dominated contraction, we have $\nu=2$. 
According to~\refeq{simi} there are two scale-invariant solutions, $(\nu,m)=(2,3)$ and $(\nu,m)=(2,0)$.
The first solution,  $(\nu,m)=(2,3)$, corresponds to the scale-invariant spectrum  observed by Wands~\cite{Wands:1998yp}. The spectrum of perturbations takes the following form,
\begin{equation}
P\propto k^{-3}\eta^{-6}~.
\end{equation}
However the magnitude of such a spectrum is amplified during the contraction process. 
The time dependence would make the power spectrum implicitly $k$-dependent at the moment of horizon-crossing and ever since. And the strongly blue-shifted power spectrum implies instabilities in the subsequent cosmological evolution.  This type of spectra often arise in the matter-dominated phase of contraction in a number of cosmological models, 
e.g.~\cite{matterbounce}, loosely called  ``matter-bounce''  models.

The second solution, $(\nu,m)=(2,0)$, is not only scale-invariant but also magnitude-invariant. 
The power spectrum in this case takes the following form:
\begin{equation}
P\, \propto k^{-3}\eta^0.
\end{equation}
which is the power spectrum of perturbations due to the tachyon field during the contracting phase of coupled-scalar-tachyon bounce cosmos (CSTBC)~\cite{Li:2011nj}. 

For the CSTB model,  in the Newtonian gauge,  $g_{\mu\nu}=diag \{-1-2\psi, a^2(1+2\psi))\delta_{ij}\}$,  the equations of motion governing the  perturbations for scalar field $\phi$ and tachyon field $T$ take the  following forms:
\begin{equation}
-\ddot{\delta\phi}+2\psi\ddot{\phi}-k^{2}a^{-2}\delta\phi+\left(-3H\dot{\delta\phi}-4\dot{\psi}\dot{\phi}+6H\psi\dot{\phi}\right)-\left(m^2_\phi+2\lambda T^2\right)\delta \phi=0~,
\end{equation}
and
\begin{eqnarray}
&\ddot{\delta T}& -\, 2\, \psi\, \ddot{T}+k^2\, \delta T\, a^{-2}\, +\, (2\, \psi\, \dot{T}^2-2\dot{\delta T}\, \dot{T})\, \left(-\frac{1}{\sqrt{2}}+3H\dot{T}+2\lambda\phi^2T \frac{\sqrt{1-\dot{T}^2}}{V(T)}\right)\\ \nonumber
&+& (1-\dot{T}^2) \left[4\, \dot\psi\,\dot{T}\, +3H\, \dot{T}\, -6\, H\, \psi\,\dot{T}\right]\\ \nonumber
&+& (1-\dot{T}^2)
\left[2\lambda\, \sqrt{1-\dot{T}^2}\, e^{T/\sqrt{2}}\, \phi\, \left(2\, T\, \delta\psi\, +\, \frac{\sqrt{2}+1}{\sqrt{2}}\phi\, \delta T + \psi\, \dot{T}^2 \, \phi -\dot{T}\, \phi\, \dot{\delta T}\right)\right]=0~;
\end{eqnarray}
where $\delta\phi$ and $\delta T$ are, respectively, the perturbations  of the scalar field, $\phi$,  and the tachyon field,  $T$.

During the  phase of tachyon matter domination in the CSTB model, the background fields $\phi$ and $T$ oscillate swiftly, see~\cite{Li:2011nj} for details~\footnote{An analytic study of the dynamics of two coupled scalar fields can be found in~\cite{Wang:2011ed}}. Such a dynamical attractor behavior ensures that
\begin{equation}
\langle\phi\rangle=\langle\dot{\phi}\rangle=\langle\ddot{\phi}\rangle=\langle\ddot{T}\rangle=\langle1-\dot{T}^2\rangle=\left\langle-\frac{1}{\sqrt{2}}+3H\dot{T}+2\lambda\phi^2T \frac{\sqrt{1-\dot{T}^2}}{V(T)}\right\rangle=0 \label{eq-azc}
\end{equation}
where $\langle A \rangle$ stands for the average of  
$A$ in a relative long period. 
Using the above conditions to simplify the general perturbation equations of the tachyon and the scalar field  we obtain
\begin{equation}
\ddot{\delta\phi}+k^{2}a^{-2}\delta\phi+3H\dot{\delta\phi}+\left(m^2_\phi+2\lambda T^2\right)\delta \phi=0 \label{eq-ppss}
\end{equation}
and
\begin{equation}
\ddot{\delta T}+\frac{k^2}{a^2}\delta T=0~.
\label{eq-ppts}
\end{equation}
The effective mass square of $\delta\phi$, $M_{eff}= m^2_\phi+2\lambda T^2$, is very large during the contracting phase with tachyon matter domination. 
Therefore, the power spectrum of scalar field perturbation is highly suppressed, and its contribution can  be neglected safely. 
With~\refeq{ppts}, we find that the effective value of $m$ for the tachyon field perturbation  vanishes due to an accidental cancelation~\refeq{azc}.  
And in the CBST model, the contracting background is also matter-dominated and hence   $\nu=2$. 
Therefore the power spectrum of tachyon field perturbations in the 
matter-dominated contracting phase is none other  but 
the solution $(\nu,m)=(2,0)$, leading to
\begin{equation}
P_{\delta T}\propto k^{-3}\eta^0,
\end{equation}
which is  both  scale-invariant and magnitude-invariant.

\section{Discussion and conclusion}
\label{sec:disc}

In this paper we study the power spectra of cosmological perturbations using their equations of motion  with a general blue/red-shift term, $mH\dot{\chi}$. The implicit time-dependence of a power spectrum  
is uncovered and its consequent influence on the scale dependence of the power spectrum  is discussed case by case.
Only if a power spectrum  is  both scale-invariant as well as magnitude-invariant (time-independent),  
can it be truly scale-invariant at all time. 
Otherwise, the power spectrum would have implicit $k$-dependence, rendering it non-scale-invariant. 

Combined the power-law index of cosmological background, $\nu$,  with the blue/red-shift parameter, $m$, 
we can construct a parameter space, $(\nu, m)$, to classify perturbation spectra. 
In the $(\nu,m)$ space,  the magnitude-invariant solutions are found to occupy a two dimensional region rather than one dimensional curves as one may have  naively expected--implying that a large portion of cosmological models could have magnitude-invariant power spectra. 
We, moreover, obtain two groups of scale-invariant solutions: one of them is magnitude-invariant 
and the other is magnitude-non-invariant. 
The number of truly scale-invariant solutions is infinite rather than preciously few--a conviction commonly held. 

In the process, we also unveil a  duality between the region of the scale-invariant solutions and the boundary of magnitude-invariant region: under two consecutive duality transformations, the scale-invariant solutions are mapped onto the boundary of magnitude invariant region. 
The physical origin of such a duality is quite worthy of further studies. 
In the last part of the paper we present a few cosmological applications of our general analysis, to a de-Sitter universe, an Ekpyrotic cosmos, and a matter-dominated contraction. 
 Previously known results confirmed our general analysis.  
 A new power spectrum that is both scale-invariant as well as magnitude-invariant is uncovered by our method in a matter-dominated contracting background. 
  This new solution to a matter-dominated contracting background was found previously in the coupled scalar-tachyon bouncing universe model~\cite{Li:2011nj}.  


One point worth re-iterated here: our result is valid for all perturbation spectra produced by scalar quantum fluctuations with the equation of motion,
$%
\ddot{\chi}_k+mH\dot{\chi}_k+\frac{k^2}{a^2}\chi_k=0~. 
$ %
For density perturbations produced by other mechanisms such as thermal fluctuations~\cite{Nayeri:2005ck, Brandenberger:2008nx}, holographic principle~\cite{Fischler:1998st, Li:2008qh, Li:2012xf} and  possibly others, e.g.~\cite{Cai:2009rd}, our analysis is, however, not applicable. 
The reason  is that this above equation of motion describing   cosmological perturbations is not the most general possible formulation of  density perturbations  produced by  a given cosmological model. 
It hence calls for a systematic way to simplify the dynamical equations of cosmological perturbations in  various gauges to the above  form.  An attempt in  this direction is underway~\cite{LiCheung2012}. 

We take the liberty to conjecture  that more cosmological models, whose the equations of motion for the cosmological perturbations have a general blue/red-shift term $mH\dot{\chi}$,  can be  constructed  in multitude in the near future.  
We also hope that this phenomenological work would perhaps give some hints from a slightly more fundamental prospective  in the search of  alternative cosmological models  with  scale-invariant power spectra that can withstand the the test of time, that is scale invariance without implicit or explicit time dependence.

\section{Acknowledgments}

Useful discussions with Yifu Cai, Konstantin Savvidy, Henry Tye and Lingfei Wang are gratefully acknowledged. This research project has been supported in parts by 985 Projects Grants from Chinese Ministry of Education,
by the Priority Academic Program Development of Jiangsu Higher Education Institutions (PAPD)
as well as by the Swedish Research Links programme under contract number 348-2008-6049 of the Swedish Research Council (Vetenskapsradets generella villkor).


\begin{thebibliography}{99}

\bibitem{Wands:1998yp} 
  D.~Wands,
  ``Duality invariance of cosmological perturbation spectra,''
  Phys.\ Rev.\ D {\bf 60}, 023507 (1999)
  [gr-qc/9809062].
   

\bibitem{Finelli:2001sr} 
  F.~Finelli and R.~Brandenberger,
  ``On the generation of a scale invariant spectrum of adiabatic fluctuations in cosmological models with a contracting phase,''
  Phys.\ Rev.\ D {\bf 65}, 103522 (2002)
  [hep-th/0112249].
  
\bibitem{Khoury:2001zk} 
  J.~Khoury, B.~A.~Ovrut, P.~J.~Steinhardt and N.~Turok,
  ``Density perturbations in the ekpyrotic scenario,''
  Phys.\ Rev.\ D {\bf 66}, 046005 (2002)
  [hep-th/0109050].\\
  P.~J.~Steinhardt and N.~Turok,
  ``The Cyclic universe: An Informal introduction,''
  Nucl.\ Phys.\ Proc.\ Suppl.\  {\bf 124}, 38 (2003)
  [astro-ph/0204479].\\
  P.~J.~Steinhardt and N.~Turok,
  ``The Cyclic model simplified,''
  New Astron.\ Rev.\  {\bf 49}, 43 (2005)
  [astro-ph/0404480].
  
  
\bibitem{Khoury:2001wf} 
  J.~Khoury, B.~A.~Ovrut, P.~J.~Steinhardt and N.~Turok,
  ``The Ekpyrotic universe: Colliding branes and the origin of the hot big bang,''
  Phys.\ Rev.\ D {\bf 64}, 123522 (2001)
  [hep-th/0103239].
  \\
  J.~Khoury, B.~A.~Ovrut, N.~Seiberg, P.~J.~Steinhardt and N.~Turok,
  ``From big crunch to big bang,''
  Phys.\ Rev.\ D {\bf 65}, 086007 (2002)
  [hep-th/0108187].
  
\bibitem{Creminelli:2004jg} 
  P.~Creminelli, A.~Nicolis and M.~Zaldarriaga,
  ``Perturbations in bouncing cosmologies: Dynamical attractor versus scale invariance,''
  Phys.\ Rev.\ D {\bf 71}, 063505 (2005)
  [hep-th/0411270].
  
\bibitem{Boyle:2004gv} 
  L.~A.~Boyle, P.~J.~Steinhardt and N.~Turok,
  ``A New duality relating density perturbations in expanding and contracting Friedmann cosmologies,''
  Phys.\ Rev.\ D {\bf 70}, 023504 (2004)
  [hep-th/0403026].
  Y.~-S.~Piao,
  ``On the dualities of primordial perturbation spectrums,''
  Phys.\ Lett.\ B {\bf 606}, 245 (2005)
  [hep-th/0404002].

\bibitem{Khoury:2009my} 
  J.~Khoury and P.~J.~Steinhardt,
  ``Adiabatic Ekpyrosis: Scale-Invariant Curvature Perturbations from a Single Scalar Field in a Contracting Universe,''
  Phys.\ Rev.\ Lett.\  {\bf 104}, 091301 (2010)
  [arXiv:0910.2230 [hep-th]].\\
  A.~Joyce and J.~Khoury,
  ``Scale Invariance via a Phase of Slow Expansion,''
  Phys.\ Rev.\ D {\bf 84}, 023508 (2011)
  [arXiv:1104.4347 [hep-th]].\\
  J.~Khoury and P.~J.~Steinhardt,
  ``Generating Scale-Invariant Perturbations from Rapidly-Evolving Equation of State,''
  Phys.\ Rev.\ D {\bf 83}, 123502 (2011)
  [arXiv:1101.3548 [hep-th]].
 
\bibitem{Garriga:1999vw}
J.~Garriga and V.~F.~Mukhanov,
``Perturbations in k-inflation,''
Phys.\ Lett.\ B {\bf 458}, 219 (1999)
[hep-th/9904176].
\bibitem{ArmendarizPicon:2000ah}
C.~Armendariz-Picon, V.~F.~Mukhanov and P.~J.~Steinhardt,
``Essentials of k essence,''
Phys.\ Rev.\ D {\bf 63}, 103510 (2001)
[astro-ph/0006373].
\bibitem{Khoury:2011ii}
J.~Khoury and P.~J.~Steinhardt,
``Generating Scale-Invariant Perturbations from Rapidly-Evolving
Equation of State,''
Phys.\ Rev.\ D {\bf 83}, 123502 (2011)
[arXiv:1101.3548 [hep-th]].


\bibitem{Zlatev:1998tr}
I.~Zlatev, L.~-M.~Wang and P.~J.~Steinhardt,
``Quintessence, cosmic coincidence, and the cosmological constant,''
Phys.\ Rev.\ Lett.\ {\bf 82}, 896 (1999)
[astro-ph/9807002].
\bibitem{Chiba:1999ka}
T.~Chiba, T.~Okabe and M.~Yamaguchi,
``Kinetically driven quintessence,''
Phys.\ Rev.\ D {\bf 62}, 023511 (2000)
[astro-ph/9912463].
\bibitem{Malquarti:2002iu}
M.~Malquarti and A.~R.~Liddle,
``Evolution of large scale perturbations in quintessence models,''
Phys.\ Rev.\ D {\bf 66}, 123506 (2002)
[astro-ph/0208562].


\bibitem{Li:2011nj} 
  C.~Li, L.~Wang and Y.~-K.~E.~Cheung,
  ``Bound to bounce: a coupled scalar-tachyon model for a smooth cyclic universe,''
  arXiv:1101.0202 [gr-qc].
  
\bibitem{Wang:2011ed} 
  L.~-F.~Wang,
  ``Preheating and locked inflation: an analytic approach towards parametric resonance,''
  JCAP {\bf 1112}, 018 (2011)
  [arXiv:1108.2608 [hep-th]].
\bibitem{Dodelson:2003ft} 
  S.~Dodelson,
  ``Modern cosmology,''
  Amsterdam, Netherlands: Academic Pr. (2003) 440 p
  
    
   
  
  
  
\bibitem{bouncereview}
  M.~Novello and S.~E.~P.~Bergliaffa,
  ``Bouncing Cosmologies,''
  Phys.\ Rept.\  {\bf 463}, 127 (2008)
  [arXiv:0802.1634 [astro-ph]].
  R.~H.~Brandenberger,
  ``The Matter Bounce Alternative to Inflationary Cosmology,''
  arXiv:1206.4196 [astro-ph.CO].
  
 \bibitem{multifield}
  D.~Langlois, S.~Renaux-Petel, D.~A.~Steer and T.~Tanaka,
  ``Primordial perturbations and non-Gaussianities in DBI and general multi-field inflation,''
  Phys.\ Rev.\ D {\bf 78}, 063523 (2008)
  [arXiv:0806.0336 [hep-th]].
  D.~Langlois and S.~Renaux-Petel,
  ``Perturbations in generalized multi-field inflation,''
  JCAP {\bf 0804}, 017 (2008)
  [arXiv:0801.1085 [hep-th]].
 
 \bibitem{matterbounce}
 Y.~F.~Cai, T.~Qiu, Y.~S.~Piao, M.~Li and X.~Zhang,
  ``Bouncing Universe with Quintom Matter,''
  JHEP {\bf 0710}, 071 (2007)
  [arXiv:0704.1090 [gr-qc]];
 Y.~F.~Cai, T.~T.~Qiu, J.~Q.~Xia and X.~Zhang,
  ``A Model Of Inflationary Cosmology Without Singularity,''
  arXiv:0808.0819 [astro-ph].
Y.~F.~Cai, T.~Qiu, R.~Brandenberger, Y.~S.~Piao and X.~Zhang,
  ``On Perturbations of Quintom Bounce,''
  JCAP {\bf 0803}, 013 (2008)
  [arXiv:0711.2187 [hep-th]];
    Y.~F.~Cai and X.~Zhang,
  ``Evolution of Metric Perturbations in Quintom Bounce model,''
  arXiv:0808.2551 [astro-ph].
  X.~Gao, Y.~Wang, R.~Brandenberger and A.~Riotto,
  ``Cosmological Perturbations in Horava-Lifshitz Gravity,''
  Phys.\ Rev.\ D {\bf 81}, 083508 (2010)
  [arXiv:0905.3821 [hep-th]];
   R.~Brandenberger,
  ``Matter Bounce in Horava-Lifshitz Cosmology,''
  Phys.\ Rev.\  D {\bf 80}, 043516 (2009)
  [arXiv:0904.2835 [hep-th]];
  Y.~-F.~Cai and E.~N.~Saridakis,
  ``Non-singular cosmology in a model of non-relativistic gravity,''
  JCAP {\bf 0910}, 020 (2009)
  [arXiv:0906.1789 [hep-th]].
  X.~Gao, Y.~Wang, W.~Xue and R.~Brandenberger,
  ``Fluctuations in a Horava-Lifshitz Bouncing Cosmology,''
  JCAP {\bf 1002}, 020 (2010)
  [arXiv:0911.3196 [hep-th]].
  Y.~-F.~Cai, D.~A.~Easson and R.~Brandenberger,
  ``Towards a Nonsingular Bouncing Cosmology,''
  JCAP {\bf 1208}, 020 (2012)
  [arXiv:1206.2382 [hep-th]].
  T.~Qiu, J.~Evslin, Y.~-F.~Cai, M.~Li and X.~Zhang,
  ``Bouncing Galileon Cosmologies,''
  JCAP {\bf 1110}, 036 (2011)
  [arXiv:1108.0593 [hep-th]].
  D.~A.~Easson, I.~Sawicki and A.~Vikman,
  ``G-Bounce,''
  JCAP {\bf 1111}, 021 (2011)
  [arXiv:1109.1047 [hep-th]].
  C.~Lin, R.~H.~Brandenberger and L.~Levasseur Perreault,
  ``A Matter Bounce By Means of Ghost Condensation,''
  JCAP {\bf 1104}, 019 (2011)
  [arXiv:1007.2654 [hep-th]].
 Y.~-F.~Cai, T.~-t.~Qiu, R.~Brandenberger and X.~-m.~Zhang,
  ``A Nonsingular Cosmology with a Scale-Invariant Spectrum of Cosmological Perturbations from Lee-Wick Theory,''
  Phys.\ Rev.\ D {\bf 80}, 023511 (2009)
  [arXiv:0810.4677 [hep-th]].
  R.~Brandenberger, H.~Firouzjahi and O.~Saremi,
  ``Cosmological Perturbations on a Bouncing Brane,''
  JCAP {\bf 0711}, 028 (2007)
  [arXiv:0707.4181 [hep-th]].
  L.~R.~Abramo and P.~Peter,
  ``K-Bounce,''
  JCAP {\bf 0709}, 001 (2007)
  [arXiv:0705.2893 [astro-ph]].
  T.~Biswas, A.~Mazumdar and W.~Siegel,
  ``Bouncing universes in string-inspired gravity,''
  JCAP {\bf 0603}, 009 (2006)
  [arXiv:hep-th/0508194].
  T.~J.~Battefeld and G.~Geshnizjani,
  ``Perturbations in a regular bouncing universe,''
  Phys.\ Rev.\ D {\bf 73}, 064013 (2006)
  [hep-th/0503160].
  J.~Martin and P.~Peter,
  ``Parametric amplification of metric fluctuations through a bouncing phase,''
  Phys.\ Rev.\ D {\bf 68}, 103517 (2003)
  [hep-th/0307077].


\bibitem{Nayeri:2005ck}
A.~Nayeri, R.~H.~Brandenberger and C.~Vafa,
``Producing a scale-invariant spectrum of perturbations in a
Hagedorn phase of string cosmology,''
Phys.\ Rev.\ Lett.\ {\bf 97}, 021302 (2006)
[hep-th/0511140].

\bibitem{Brandenberger:2008nx}
R.~H.~Brandenberger,
``String Gas Cosmology,''
arXiv:0808.0746 [hep-th].

\bibitem{Fischler:1998st}
W.~Fischler and L.~Susskind,
``Holography and cosmology,''
hep-th/9806039.
\bibitem{Li:2008qh}
M.~Li, X.~-D.~Li, C.~Lin and Y.~Wang,
``Holographic Gas as Dark Energy,''
Commun.\ Theor.\ Phys.\ {\bf 51}, 181 (2009)
[arXiv:0811.3332 [hep-th]].
\bibitem{Li:2012xf}
M.~Li and R.~-X.~Miao,
``A New Model of Holographic Dark Energy with Action Principle,''
arXiv:1210.0966 [hep-th].
 
\bibitem{Cai:2009rd} 
  Y.~-F.~Cai, W.~Xue, R.~Brandenberger and X.~-m.~Zhang,
  ``Thermal Fluctuations and Bouncing Cosmologies,''
  JCAP {\bf 0906}, 037 (2009)
  [arXiv:0903.4938 [hep-th]].

\bibitem{LiCheung2012} 
C.~ Li and Y.~K.~E.~Cheung, in preparation.


\end{thebibliography}
\end{document}